\documentclass[conf]{new-aiaa}
\usepackage[utf8]{inputenc}

\usepackage{graphicx}
\usepackage{amsmath}
\usepackage[version=4]{mhchem}
\usepackage{siunitx}
\usepackage{longtable,tabularx}
\title{Development and Validation of a Comprehensive Helicopter Flight Dynamics Code}

\author{Bochan Lee \footnote{Graduate Research Assistant, Aerospace Engineering, Texas A\&M University, College Station, TX 77843} and Moble Benedict \footnote{Assistant Professor, Aerospace Engineering, Texas A\&M University, College Station, TX 77843}}
\affil{Aerospace Engineering, Texas A\&M University, College Station, TX, 77843}

\begin{document}

\maketitle

\begin{abstract}
A comprehensive helicopter flight dynamics code is developed based on the UH-60 helicopter and named Texas A\&M University Rotorcraft Analysis Code (TRAC). This is a complete software package, which could perform trim analysis to autonomous flight simulation and the capability to model any helicopter configuration. Different components of the helicopter such as the main rotor, tail rotor, fuselage, vertical tail, and horizontal tail are modeled individually as different modules in the code and integrated to develop a complete UH-60 model. Since the code is developed on a module basis, it can be easily modified to adopt another component or configure a different helicopter. TRAC can predict the dynamic responses of both the articulated rotor blades and the helicopter fuselage and yields the required pilot control inputs to achieve trim condition for different flight regimes such as hover, forward flight, coordinated turn, climb/descent, etc. These trim results are validated with the test data obtained from the UH-60 flight tests conducted by the US Army. Beyond trim analysis, TRAC can also generate linearized models at various flight conditions based on a first-order Taylor series expansion. The extracted linear models show realistic helicopter dynamic behavior and were used to simulate a fully autonomous flight that involves a UH-60 helicopter approaching a ship and landing on the deck by implementing a Linear Quadratic Regulator (LQR) optimal controller.

\end{abstract}

\section{Nomenclature}

{\renewcommand\arraystretch{1.0}
\noindent\begin{longtable*}{@{}l @{\quad=\quad} l@{}}
$c$ & chord length\\
$c_{l}$, $c_{d}$, $c_{m}$ & Blade section lift, drag and pitching moment coefficients\\
$C_{T}$, $C_{T,TR}$, $C_{Q,TR}$ & main rotor lift coefficient, tail rotor lift coefficient, tail rotor torque coefficient \\
$dr$ & blade element of span\\
$dD$ & blade elemental drag\\
$dL$ & blade elemental lift\\
$eR$ & hinge offset distance from hub\\
$F_{MR,I}$ & main rotor inertial forces\\
$i_{B}$, $j_{B}$, $k_{B}$ & Unit vectors of body-fixed frame\\
$i_{L}$, $j_{L}$, $k_{L}$ & Unit vectors of blade lagged frame\\ 
$m_{b}$ & main rotor blade mass per unit length\\
$M_{MR,I}$ & main rotor inertial moments\\ 
$M_{FLAP,A}$ & aerodynamic flap moment \\
$N_{b}$ & number of blades\\
$p_{F}$, $q_{F}$, $r_{F}$ & angular velocities about 3 body-fixed frames, rad/s \\
$Q_{TR}$ & tail rotor torque\\
$R$, $R_{TR}$ & main rotor blade radius, tail rotor blade radius\\
$T_{TR}$ & tail rotor thrust\\
$u_{F}$, $v_{F}$, $w_{F}$ & linear velocities along 3 body-fixed frames, m/s \\
$y_F$  & airframe rigid-body states \\
$y_\lambda$ & induced inflow coefficients for main rotor and tail rotor \\
$y_{rotor}$ & rotor deflection states \\
$\alpha$ & effective angle of attack\\
$\beta_{1c}$ & nose-down tilt of the rotor tip path plane \\
$\theta$ & blade pitch angle\\
$\theta_{o}$ & collective pitch\\
$\theta_{1c}$ & lateral cyclic \\
$\theta_{1s}$ & longitudinal cyclic\\
$\theta_{tw}$ & blade twist\\
$\theta_{H}$ & pitch of horizontal tail (stabilator) \\
$\lambda_{o}$, $\lambda_{TR}$ & main rotor uniform inflow ratio, tail rotor uniform inflow ratio\\
$\lambda_{1c}$ & main rotor sine inflow ratio\\
$\lambda_{1s}$ & main rotor cosine inflow ratio\\
$\mu$ & advance ratio  \\
$\rho$ & air density\\
$\phi$ & relative wind angle\\
$\phi_{F}$, $\theta_{F}$, $\psi_{F}$ & fuselage Euler angles about earth-fixed frame\\ 
$\chi$ & wake skew angle \\
$\Omega_{MR}$, $\Omega_{TR}$ & main rotor rotational speed, tail rotor rotational speed\\

\end{longtable*}}

\section{Introduction} 
\lettrine{I}{n} the last few years, with the increased demand for advanced aircraft such as tiltrotors, compound helicopters (helicopters with thrusters and wings), and Vertical Take-Off and Landing (VTOL) Unmanned Aerial Vehicles (UAVs), there is a need to develop modular analysis tools that can predict performance of different rotorcraft configurations during the early stages of development. Over the past few decades, there have been many studies to enhance the accuracy of helicopter dynamic models with the help of advances in computation methods \cite{theodore2000helicopter,ribera2007helicopter,sridharan2014simulation}. These previous studies mainly focus on the improvement of helicopter trim power predictions by adding a free wake model. The helicopter dynamic models coupled with the free wake model will increase the computational load; however, it improves the accuracy of power prediction at lower forward flight speeds. In the present study, a comprehensive helicopter dynamics mathematical model is developed focusing on its functionality in aircraft design and evaluation. This framework is named Texas A\&M University Rotorcraft Analysis Code (TRAC). The aerodynamic/dynamic models of the individual helicopter components are developed separately and integrated into a complete helicopter model. Because of the modularity of the framework, any of the components could be easily replaced with a new component model, which enables TRAC to analyze how changes in the component affect the performance of the helicopter. In this way, it can be used for verifying a new component or configuring a different helicopter. For example, TRAC is currently being used in a different study to investigate the benefits of a mission-adaptive morphing rotor \cite{allen2020Biomimetic}. Although the present modeling methodology does not utilize a free wake model in order to reduce computational time, it achieves good accuracy by adopting wind-tunnel tests and empirical data to multiple component dynamic models. \\
\indent A UH-60 helicopter is selected as a baseline model due to the ample amount of flight test data available for validation purposes. The fuselage, main rotor, tail rotor, horizontal tail (stabilator), and vertical tail (fin) models are individually developed and integrated into a complete UH-60 model. This fully non-linear mathematical UH-60 helicopter model yields trim results for helicopter attitude, rotor blade flap/lag angles, power, and control inputs at specified forward velocities and are validated with the US Army flight test data \cite{abbott1982validation}. The rotor model uses Pitt-Peters linear inflow model \cite{pitt1980theoretical,peters1988dynamic} to capture the main rotor wake, and the predictions show good agreements with flight test data. The trim results, which include the helicopter dynamic responses and control inputs are presented in this paper.\\ 
\indent Beyond the trim analysis, TRAC can also extract linearized models at various flight conditions such as hover, vertical ascent and descent, level flight, ascent and descent with forward velocity, and coordinated turn with or without ascent and descent. All of these linear models are extracted based on the first-order Taylor series expansion and these computed models can be used for any linear control system. In the present study, the Linear Quadratic Regulator (LQR) optimal controller is applied to stabilize the helicopter responses and achieve a fully autonomous flight using a trajectory tracking method. The helicopter model along with the LQR controller is implemented in MATLAB and the results are visualized using the X-Plane flight simulator. In order to demonstrate the capability of TRAC, a complicated helicopter maneuver, which includes ship approach and landing, is simulated as an example. It consists of descent, steady forward flight, steady coordinated turn, deceleration, and final landing. For each phase of this complex flight maneuver, different linearized models for the UH-60 helicopter dynamics are extracted. The simulated flight spans more than 0.5 nautical miles; however, it shows only a few centimeters of error for the final landing position. The simulation also verifies the realistic helicopter dynamic behavior including pilot control inputs throughout the entire maneuver. \\
\indent TRAC has been developed as a complete helicopter flight dynamics tool, which includes procedures from modeling to control systems. Inevitably, the modeling methodology has adopted many things from the previous methods; however, TRAC has been developed with modularity and computational speed in mind so that it can be used for quick performance evaluation at the conceptual design stage. 

\section{Modeling Methodology}
The helicopter is modeled as a rigid body with rotating articulated blades that can undergo flap, lag and pitch motions about hinges. The UH-60 helicopter model is based on Blade Element Momentum Theory (BEMT) and consists of dynamic and aerodynamic models of the fuselage, main rotor, tail rotor and empennage (vertical tail and horizontal tail). The UH-60 parameters used for the modeling are specified in \textbf{Table. \ref{config table}}.   
The governing equations of the system are formulated in state-space form as a system of first-order nonlinear coupled ODEs :
\begin{equation}
\label{first-order ODEs}
f(y, \hspace{0.3cm} \dot{y}, \hspace{0.3cm} u, \hspace{0.3cm} t) \hspace{0.3cm} = \hspace{0.3cm} \epsilon \hspace{0.3cm} = \hspace{0.3cm} 0
\end{equation}
\begin{equation} 
\label{system state vector}
y \hspace{0.3cm} = \hspace{0.3cm} \{\hspace{0.3cm}  y_F^T \hspace{0.3cm} y_\lambda^T \hspace{0.3cm} y_{rotor}^T \hspace{0.3cm} \}^T
\end{equation}
\begin{equation} 
\label{control state vector}
u \hspace{0.3cm} = \hspace{0.3cm} \{\hspace{0.3cm} \delta_{col} \hspace{0.3cm} \delta_{lat} \hspace{0.3cm} \delta_{lon} \hspace{0.3cm} \delta_{ped} \hspace{0.3cm}\}^T
\end{equation}
y is a vector of system states, $\dot{y}$ is time derivatives of y, u is a vector of control inputs, and t is the time. Numerical solutions of these equations with zero body-axis accelerations for trim are used to study vehicle performance in steady flight. 

\subsection{Fuselage}
The helicopter fuselage is modeled as a rigid body, and the inertial loads can be computed from the body-axis components of the airframe linear and angular velocities. These components are obtained from the partition of the system state vector that contains the fuselage states, given by
\begin{equation}
\label{fuselage states}
y_{F} \hspace{0.3cm} = \hspace{0.3cm} \{ \hspace{0.3cm} u_{F} \hspace{0.3cm} v_{F} \hspace{0.3cm} w_{F} \hspace{0.3cm} p_{F} \hspace{0.3cm} q_{F} \hspace{0.3cm} r_{F} \hspace{0.3cm} \phi_{F} \hspace{0.3cm} \theta_{F} \hspace{0.3cm} \psi_{F} \hspace{0.3cm} \}^{T}
\end{equation}
Since the fuselage is rigid, the position and orientation of the force-producing components (main rotor, tail rotor, horizontal and vertical tail) remain constant in the body-fixed frame and the moments of inertia of a rigid body stay the same in the body-fixed frame. Hence, it is convenient to formulate force and moment equilibrium equations about the helicopter fuselage body-fixed frame. The force equilibrium equations are
\begin{equation}
\label{force equilibrium}
\begin{aligned}
X \hspace{0.3cm} &= \hspace{0.3cm} m_{F}(\dot{u_{F}} + q_{F}w_{F} - r_{F}v_{F} + g\sin{\theta_{F}}) \\ 
Y \hspace{0.3cm} &= \hspace{0.3cm} m_{F}(\dot{v_{F}} + r_{F}u_{F} - p_{F}w_{F} - g\sin{\phi_{F}}\cos{\theta_{F}}) \\
Z \hspace{0.3cm} &= \hspace{0.3cm} m_{F}(\dot{w_{F}} + p_{F}v_{F} - q_{F}u_{F} - g\cos{\phi_{F}}\cos{\theta_{F}})
\end{aligned}
\end{equation}
The moment equilibrium equations are
\begin{equation}
\label{moment equilibrium}
\begin{aligned}
L \hspace{0.3cm} &= \hspace{0.3cm} I_{xx}\dot{p_{F}} - I_{xy}(\dot{q_{F}} - p_{F}r_{F}) - I_{xz}(\dot{r_{F}} - p_{F}q_{F}) - I_{yz}(q_{F}^{2} - r_{F}^{2}) - (I_{yy} - I_{zz})q_{F}r_{F}\\
M \hspace{0.3cm} &= \hspace{0.3cm} I_{yy}\dot{q_{F}} - I_{yz}(\dot{r_{F}} - q_{F}p_{F}) - I_{yx}(\dot{p_{F}} - q_{F}r_{F}) - I_{zx}(r_{F}^{2} - p_{F}^{2}) - (I_{zz} - I_{xx})r_{F}p_{F}\\
N \hspace{0.3cm} &= \hspace{0.3cm} I_{zz}\dot{r_{F}} - I_{zx}(\dot{p_{F}} - r_{F}q_{F}) - I_{zy}(\dot{q_{F}} - r_{F}p_{F}) - I_{xy}(p_{F}^{2} - q_{F}^{2}) - (I_{xx} - I_{yy})p_{F}q_{F}
\end{aligned}
\end{equation}
The terms on the left-hand side of \textbf{Eqs. (\ref{force equilibrium}) and (\ref{moment equilibrium})} represent the cumulative forces and moments about the center of gravity, respectively. These are exerted by airframe aerodynamics, main rotor loads, tail rotor loads, and empennage aerodynamics, and given by
\begin{equation}
\label{external force and moment}
\begin{aligned}
X \hspace{0.3cm} &= \hspace{0.3cm} X_{MR} \hspace{0.3cm} + \hspace{0.3cm} X_{TR} \hspace{0.3cm} + \hspace{0.3cm} X_{H} \hspace{0.3cm} + \hspace{0.3cm} X_{V} \hspace{0.3cm} + \hspace{0.3cm} X_{F} \\
Y \hspace{0.3cm} &= \hspace{0.3cm} Y_{MR} \hspace{0.3cm} + \hspace{0.3cm} Y_{TR} \hspace{0.3cm} + \hspace{0.3cm} Y_{H} \hspace{0.3cm} + \hspace{0.3cm} Y_{V} \hspace{0.3cm} + \hspace{0.3cm} Y_{F}\\
Z \hspace{0.3cm} &= \hspace{0.3cm} Z_{MR} \hspace{0.3cm} + \hspace{0.3cm} Z_{TR} \hspace{0.3cm} + \hspace{0.3cm} Z_{H} \hspace{0.3cm} + \hspace{0.3cm} Z_{V} \hspace{0.3cm} + \hspace{0.3cm} Z_{F}\\
L \hspace{0.3cm} &= \hspace{0.3cm} L_{MR} \hspace{0.3cm} + \hspace{0.3cm} L_{TR} \hspace{0.3cm} + \hspace{0.3cm} L_{H} \hspace{0.3cm} + \hspace{0.3cm} L_{V} \hspace{0.3cm} + \hspace{0.3cm} L_{F}\\
M \hspace{0.3cm} &= \hspace{0.3cm} M_{MR} \hspace{0.3cm} + \hspace{0.3cm} M_{TR} \hspace{0.3cm} + \hspace{0.3cm} M_{H} \hspace{0.3cm} + \hspace{0.3cm} M_{V} \hspace{0.3cm} + \hspace{0.3cm} M_{F}\\
N \hspace{0.3cm} &= \hspace{0.3cm} N_{MR} \hspace{0.3cm} + \hspace{0.3cm} N_{TR} \hspace{0.3cm} + \hspace{0.3cm} N_{H} \hspace{0.3cm} + \hspace{0.3cm} N_{V} \hspace{0.3cm} + \hspace{0.3cm} N_{F}
\end{aligned}
\end{equation}
Fuselage drag force is the dominant component in the fuselage aerodynamic force. There have been studies to estimate the equivalent flat-plate area for a helicopter fuselage. In these studies, models to estimate the flat-plate area as a function of the fuselage angle of attack has been developed through wind-tunnel testing. To predict the fuselage drag force precisely, several different estimations were tested and the model finally selected for the present study is the estimation by Yeo et al. \cite{yeo2004performance} and given in \textbf{Eq. (\ref{fuselage drag})}, where 35.14 $ft^{2}$ is the flat-plate area of the UH-60 helicopter fuselage at zero angle of attack. The fuselage drag force is computed by multiplying the flat-plate area with dynamic pressure. 
\begin{equation}
\label{fuselage drag}
\begin{aligned}
D_{F} \hspace{0.3cm} &= \hspace{0.3cm} \frac{1}{2} \rho {u_{F}}^2 f(\alpha_{F})\\
f(\alpha_{F}) \hspace{0.3cm} &= \hspace{0.3cm} 35.14 \hspace{0.2cm} + \hspace{0.2cm} 0.016(1.66 \alpha_{F})^2
\end{aligned}
\end{equation}
Since the drag direction is parallel to the wind direction, the fuselage has force contribution along the body-fixed x (forward) and z (downward) direction.

\subsection{Main Rotor}

\begin{figure}[ht!]
\vspace{0cm}
\centering
	\includegraphics[scale=0.3]{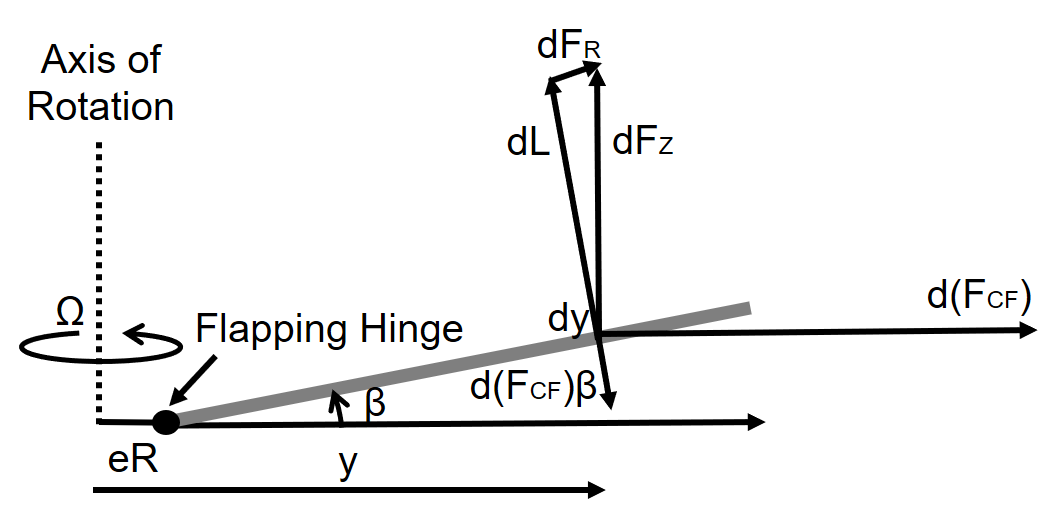}
	\caption{Schematic Showing Blade Flapping Equilibrium}
	\label{flapping schematic}
\end{figure}

\begin{figure}[ht!]
\vspace*{0cm}
\centering
	\includegraphics[scale=0.3]{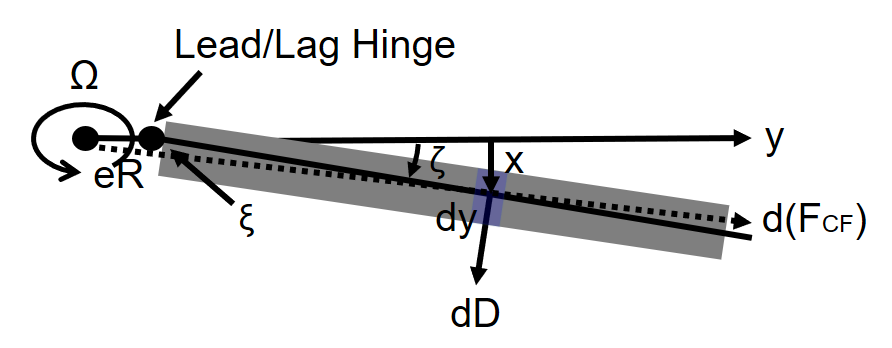}
	\caption{Schematic Showing Blade Lagging Equilibrium}
	\label{lagging schematic}
\end{figure}

The UH-60 helicopter main rotor has four blades and each blade experiences flap and lead-lag angular motions which are given by $\beta$ and $\zeta$, respectively, shown in \textbf{Fig. \ref{flapping schematic} and \ref{lagging schematic}}. The flap and lead-lag hinge with its own spring and damper are placed at the same location. The blade equations of motion are nonlinear, coupled, partial differential equations with periodic coefficients. In this paper, the inertial and aerodynamic load vectors are calculated numerically with the assumption of the first harmonic blade motion. Each blade is discretized spatially with 100 blade elements in the spanwise direction and temporally with $1^{\circ}$ increments in azimuth. This results in a total of 36,000 data points for one blade in one revolution to calculate lift and drag. In addition, the negative twist angle of $18^{\circ}$ is considered in the computation of the elemental angle of attack. The summation of the loads on each blade element is collected and transformed into the body-fixed frame.

\begin{figure}[hbt!]
\centering
   \begin{minipage}{0.48\textwidth}
     \centering
     \includegraphics[scale=0.3]{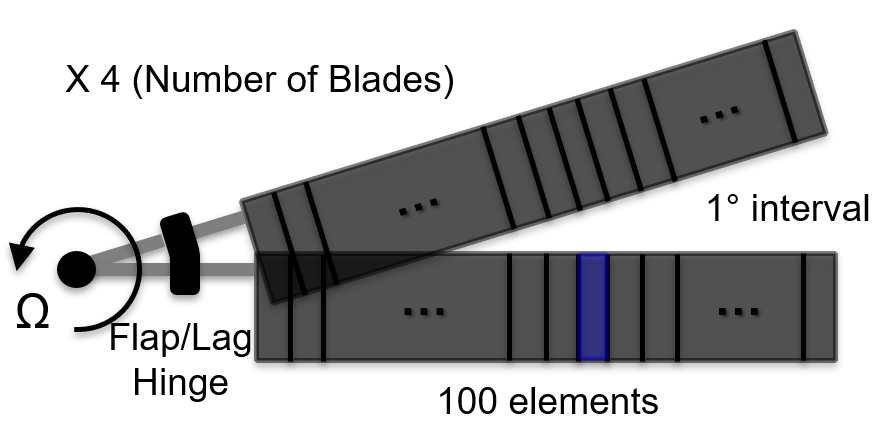}
     \caption{Discretized blade elements}
   \end{minipage}
   \begin {minipage}{0.48\textwidth}
    \centering
	\includegraphics[scale=0.3]{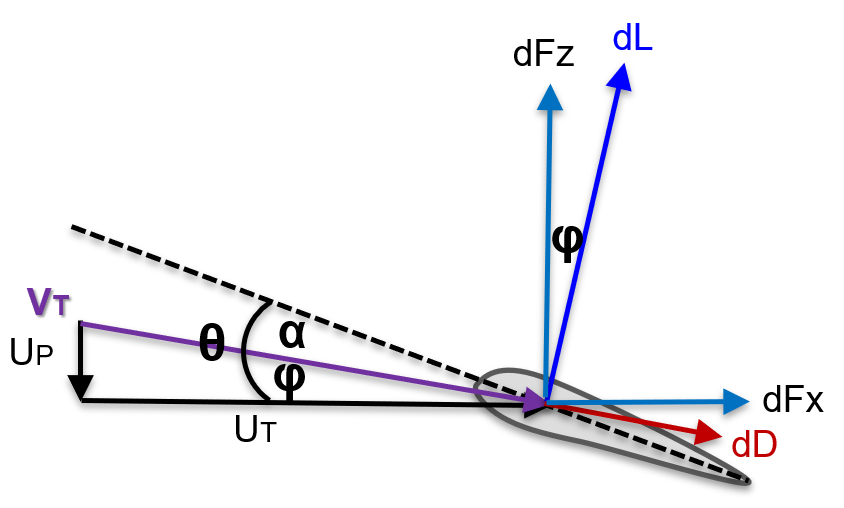}
	\caption{Blade elemental lift and drag}
   \end{minipage}
\end{figure}

In the formulation of the main rotor equations of motion, the distributed loads due to blade inertia are required. These inertia loads depend on the absolute acceleration of a point on the rotor blade, $A_{P}$. Main rotor inertial forces and moments in the body-fixed frame are obtained by integrating along the blade and along the azimuth as
\begin{equation}
\begin{aligned}
F_{MR,I} \hspace{0.5cm} &= \hspace{0.5cm} \frac{Nb}{2\pi} \int_{0}^{2\pi} \int_{eR}^{R} \hspace{0.3cm} m_{b} \hspace{0.3cm} A_{P,B} \hspace{0.3cm} dr d\psi\\
M_{MR,I} \hspace{0.5cm} &= \hspace{0.5cm} \int_{eR}^{R} \hspace{0.3cm} m_{b} \hspace{0.3cm} (R_{b,L} \hspace{0.3cm} \times \hspace{0.3cm} A_{P,L}) \hspace{0.3cm} dr
\end{aligned}
\end{equation}

$V_{T}$ is the resultant velocity of the airflow at the quarter-chord location and can be expressed as 
\begin{equation}
\label{velocity}
\begin{aligned}
    V_{T} = V_{P} - V_{I}
\end{aligned}
\end{equation}

$V_{P}$ is the velocity caused by forward flight and blade flapping. $V_{I}$ is the velocity induced at the point on the blade by the rotor wake. The resultant velocity $V_{T}$ is parallel to elemental drag $dD$ and perpendicular to elemental lift $dL$. Blade pitch angle $\theta$ is the sum of inflow relative wind angle $\phi$ and effective angle of attack $\alpha$. The elemental drag and lift are calculated by multiplying dynamic pressure, area, and coefficients. Blade section lift coefficient $c_{l}$ and drag coefficient $c_{d}$ are selected from data look-up tables as a function of local angle of attack and Mach number.  

\begin{equation}
\label{elemental force and drag}
\begin{aligned}
        dL \hspace{0.5cm}&= \hspace{0.5cm}\frac{1}{2}\hspace{0.2cm}\rho\hspace{0.2cm}V_{T}^{2}\hspace{0.2cm}c_{l}\hspace{0.2cm}c\hspace{0.2cm}dr\\
        dD \hspace{0.5cm}&= \hspace{0.5cm}\frac{1}{2}\hspace{0.2cm}\rho\hspace{0.2cm}V_{T}^{2}\hspace{0.2cm}c_{d}\hspace{0.2cm}c\hspace{0.2cm}dr\\
        dF \hspace{0.5cm}&=\hspace{0.5cm} (dL\cos{\phi} - dDsin_{\phi})\hspace{0.1cm}k_{L}\hspace{0.2cm} - \hspace{0.2cm} (dL\sin{\phi} + dD\cos{\phi})\hspace{0.1cm}j_{L} \hspace{0.2cm} 
\end{aligned}
\end{equation}
Integrating over the blade span and along the azimuth yields the total aerodynamic forces as
\begin{equation}
\label{MR aerodynamic force}
\begin{aligned}
F_{MR,A} \hspace{0.5cm} &= \hspace{0.5cm} \frac{Nb}{2\pi} \int_{0}^{2\pi} \int_{eR}^{R} \hspace{0.3cm} dF \hspace{0.3cm} dr d\psi\\
\end{aligned}
\end{equation}
Aerodynamic flap moment $M_{FLAP,A}$ about the hinge is computed by
\begin{equation}
\label{MR aerodynamic moment}
\begin{aligned}
M_{FLAP,A} \hspace{0.5cm}&=\hspace{0.5cm} \int_{eR}^{R} r \hspace{0.2cm} \times \hspace{0.2cm} dL \hspace{0.2cm} dr\\
&= \hspace{0.5cm} \frac{1}{2}\hspace{0.2cm} \rho\hspace{0.2cm} a \hspace{0.2cm}\int_{eR}^{R} \hspace{0.2cm}(\theta \hspace{0.2cm}U_{T}^2 \hspace{0.2cm}-\hspace{0.2cm} U_{P}\hspace{0.2cm}U_{T})\hspace{0.2cm} c\hspace{0.2cm} dr
\end{aligned}
\end{equation}

The equilibrium position of the blade is determined by the balance of inertial, aerodynamic, and centrifugal forces (CF). The flapping angle may be assumed small due to the fact that the centrifugal force is significantly greater than the aerodynamic force. Moment equilibrium about the flapping hinge is expressed as  
\begin{equation}
\label{flapping equilibrium}
\begin{aligned}
M_{MR,I}\hspace{0.8cm} +&\hspace{0.8cm} M_{MR,CF}\hspace{0.8cm} + \hspace{0.8cm}M_{FLAP,A}\hspace{0.2cm} =\hspace{0.2cm} 0\\
\int_{eR}^{R} m_{b}(y - eR)^{2}\ddot{\beta}dy \hspace{0.2cm} +& \hspace{0.2cm} \int_{eR}^{R} m_{b}\Omega^{2}y(y - eR)\beta dy \hspace{0.2cm} - \hspace{0.2cm} \int_{eR}^{R} L(y - eR)dy \hspace{0.2cm} = \hspace{0.2cm} 0
\end{aligned} 
\end{equation}

By replacing flapping terms to constant and periodic terms on both sides of the flapping equation, flapping angles can be related to control angles ($\theta_{o}$, $\theta_{1c}$, $\theta_{1s}$).

The equilibrium of the blade about the lead-lag hinge is determined by a balance of centrifugal and aerodynamic moments. The aerodynamic moments are generated by the aerodynamic drag of the blade as it rotates. Moment equilibrium about the lead-lag hinge is expressed as  
\begin{equation}
\label{lead-lag equilibrium}
\begin{aligned}
    M_{MR,I}\hspace{0.8cm} +&\hspace{0.8cm} M_{MR,CF}\hspace{0.8cm} + \hspace{0.8cm}M_{LAG,A}\hspace{0.2cm} =\hspace{0.2cm} 0\\
    -\int_{eR}^{R} m_{b}(y - eR)^{2}\ddot{\zeta}dy \hspace{0.2cm} +& \hspace{0.2cm} \int_{eR}^{R} m_{b}\Omega^{2}y(y - eR)\frac{eR}{y}\zeta dy \hspace{0.2cm} + \hspace{0.2cm} \int_{eR}^{R} D(y - eR)dy \hspace{0.2cm} = \hspace{0.2cm} 0
\end{aligned} 
\end{equation}
\pagebreak
 \\

The centrifugal restoring moment about the lag hinge is much smaller than in flapping, and the corresponding uncoupled natural frequency of the lag motion is much smaller. For articulated rotors such as UH-60 rotors, the uncoupled rotating lag frequency varies from about 0.2 to 0.3$\Omega$. The lead-lag displacements about the hinge are small and aerodynamic forces are produced by changes in velocity and dynamic pressure normal to the leading edge of the blade. However, it is much smaller than the aerodynamic forces which are produced through flapping motion by changes in the angle of attack. Furthermore, the drag forces acting on the blades are also much smaller than the lift forces.  
It is assumed that the main rotor inflow distribution is linear, thus it is expressed with respect to the blade azimuth $\psi_{n}$ as
\begin{equation}
\label{inflow basic}
\begin{aligned}
\lambda \hspace{0.5cm} = \hspace{0.5cm} \lambda_{o} \hspace{0.2cm} + \hspace{0.2cm} \lambda_{1c} \frac{r}{R} \cos{\psi_{n}}  \hspace{0.2cm} + \hspace{0.2cm} \lambda_{1s} \frac{r}{R} \sin{\psi_{n}} 
\end{aligned}
\end{equation}

Basically, modeling a linear inflow distribution is to estimate the values of $\lambda_{o}$, $\lambda_{1c}$, and $\lambda_{1s}$. For this mathematical UH-60 helicopter model, the Pitt-Peters linear inflow model is used. These dynamic inflow components are related to the forces on the rotor disk which are the rotor thrust, pitching moment, and rolling moment coefficient. For dynamic analysis of the blade, the dynamic inflow components are treated as additional degrees of freedom. It is formulated on the basis of experimental results or more advanced vortex theories and it is well suited for helicopter rotor aerodynamics, and flight dynamics.

\subsection{Tail Rotor}
The primary role of the tail rotor is to counter the torque effect created by the main rotor. In addition, the UH-60 helicopter tail rotor also has a small contribution to the thrust since it has 20$^{\circ}$ of cant angle. The tail rotor model is based on a simplified implementation of the closed-form solution given by F. J. Bailey \cite{bailey1941simplified}, which relates the free-stream velocity to the rotor thrust, torque, and induced inflow. The velocity at the tail rotor reference point (hub) is
\begin{equation}
\label{tail rotor induced velocity}
\begin{aligned}
    V_{TR} \hspace{0.5cm} = \hspace{0.5cm} V_{b} \hspace{0.5cm} + \hspace{0.5cm} \omega \times r_{TR} \hspace{0.5cm} + \hspace{0.5cm} V_{TR,in}
\end{aligned}
\end{equation}
$V_{TR,in}$ represents the induced velocity at the tail rotor reference point due by the wake of the main rotor and fuselage, given by
\begin{equation}
\label{tail rotor induced velocity 2}
\begin{aligned}
    V_{TR,in} \hspace{0.5cm} = \hspace{0.5cm} \lambda_{o}\Omega_{MR}R \Big[v_{x_{TR}}(\beta_{1c},\chi) i_{B} \hspace{0.5cm} + \hspace{0.5cm} v_{z_{TR}}(\beta_{1c},\chi) k_{B} \Big]
\end{aligned}
\end{equation}
where $\lambda_{o}\Omega_{MR}R$ is average main rotor downwash. The functions $v_{x,TR}$, $v_{z,TR}$ are obtained from lookup tables based on the wake skew angle $\chi$ and the tip-path plane tilt $\beta_{1c}$ with respect to the fuselage. The velocity $V_{TR}$ at the tail rotor reference point $r_{TR}$ is resolved into components along the tail rotor axes. 

\begin{equation}
\label{tail rotor force components}
\begin{aligned}
T_{TR} \hspace{0.5cm} &= \hspace{0.5cm} \rho \hspace{0.2cm} \pi \hspace{0.2cm} \Omega_{TR}^{2} \hspace{0.2cm} R_{TR}^{4} \hspace{0.2cm} C_{T,TR}\\
K_{TR} \hspace{0.5cm} &= \hspace{0.5cm} T_{TR} \hspace{0.2cm} \sin{20^{\circ}}\\
J_{TR} \hspace{0.5cm} &= \hspace{0.5cm} T_{TR} \hspace{0.2cm} \cos{20^{\circ}}
\end{aligned}
\end{equation}
$T_{TR}$ is the tail rotor thrust, which is assumed to act along the shaft direction. $K_{TR}$ and $J_{TR}$ are force components along the vertical and lateral direction respectively. The tail rotor torque due to induced and profile drag is
\begin{equation}
\label{tail rotor torque}
\begin{aligned}
Q_{TR} \hspace{0.5cm} &= \hspace{0.5cm} \rho \hspace{0.2cm} \pi \hspace{0.2cm} \Omega_{TR}^{2} \hspace{0.2cm} R_{TR}^{5} \hspace{0.2cm} C_{Q,TR} \nonumber
\end{aligned}
\end{equation}

The induced inflow of the tail rotor is assumed to be uniform over the disk and is represented using a 1-state Pitt-Peters dynamic inflow model. The ODE governing the inflow dynamics is
\begin{equation}
\label{tail rotor induced}
\begin{aligned}
\frac{4R_{TR}}{2\pi|V_{TR}|} \dot{\lambda}_{TR} \hspace{0.5cm} + \hspace{0.5cm} \lambda_{TR} \hspace{0.5cm} &= \hspace{0.5cm} \frac{C_{T,TR}\Omega_{TR}R_{TR}}{2|V_{TR}|}
\end{aligned}
\end{equation}

\subsection{Empennage}
Horizontal tail (stabilator) and vertical tail (fin) are modeled to include their functionalities for pitch and yaw of the helicopter. Horizontal tail changes its incidence angle in relation to the forward velocity to reduce fuselage nose-up motion at low airspeed and the vertical tail has a fixed angle to provide the counter-torque at high airspeeds. In order to compute the aerodynamic loads acting on the horizontal tail and vertical tail, the velocity at the reference point of each lifting surface is calculated by the fuselage translation velocity $V_{b}$ and angular velocity $w_{b}$ with the position of the reference points with respect to the vehicle center of gravity $r_{H}$, $r_{V}$.

\begin{equation}
\label{empennage velocity}
\begin{aligned}
V_{H} \hspace{0.5cm} &= \hspace{0.5cm} K_{H}V_{b} \hspace{0.5cm} + \hspace{0.5cm} \omega \times r_{H} \hspace{0.5cm} + \hspace{0.5cm} V_{H,in}\\
V_{V} \hspace{0.5cm} &= \hspace{0.5cm} K_{V}V_{b} \hspace{0.5cm} + \hspace{0.5cm} \omega \times r_{V} \hspace{0.5cm} + \hspace{0.5cm} V_{V,in}
\end{aligned}
\end{equation}
$K_{H}$ and $K_{V}$ are empirical correction factors for the dynamic pressure loss at the tail surfaces due to the airframe wake. $V_{H,in}$ and $V_{V,in}$ are the induced velocities at the tail surfaces by the main rotor wake and they are obtained from wind-tunnel tests.
\begin{equation}
\label{empennage induced}
\begin{aligned}
    V_{H,in} \hspace{0.5cm} = \hspace{0.5cm} \lambda_{o}\Omega_{MR}R \Big[v_{x_{H}}(\beta_{1c},\chi) i_{B} \hspace{0.5cm} + \hspace{0.5cm} v_{z_{H}}(\beta_{1c},\chi) k_{B} \Big]\\
    V_{V,in} \hspace{0.5cm} = \hspace{0.5cm} \lambda_{o}\Omega_{MR}R \Big[v_{x_{V}}(\beta_{1c},\chi) i_{B} \hspace{0.5cm} + \hspace{0.5cm} v_{z_{V}}(\beta_{1c},\chi) k_{B} \Big]
\end{aligned}
\end{equation}

The functions $v_{x_{H}}$, $v_{z_{H}}$, $v_{x_{V}}$, $v_{z_{V}}$ are obtained from lookup tables based on the wake skew angle $\chi$ and the tip-path plane tilt $\beta_{1c}$ with respect to the fuselage.
The pitch of the horizontal tail (stabilator) $\theta_{H}$ is scheduled to change with the fuselage speed in a prescribed manner. Using the incidence angles $\alpha$ and $\beta$ for each surface and the dynamic pressure at the reference points, the aerodynamic lift and drag coefficients are obtained using lookup table data based on wind-tunnel experiments and transformed into the helicopter body axes. 
\section{Validation}
\begin{figure}[hbt!]
\centering
	\includegraphics[scale=0.45]{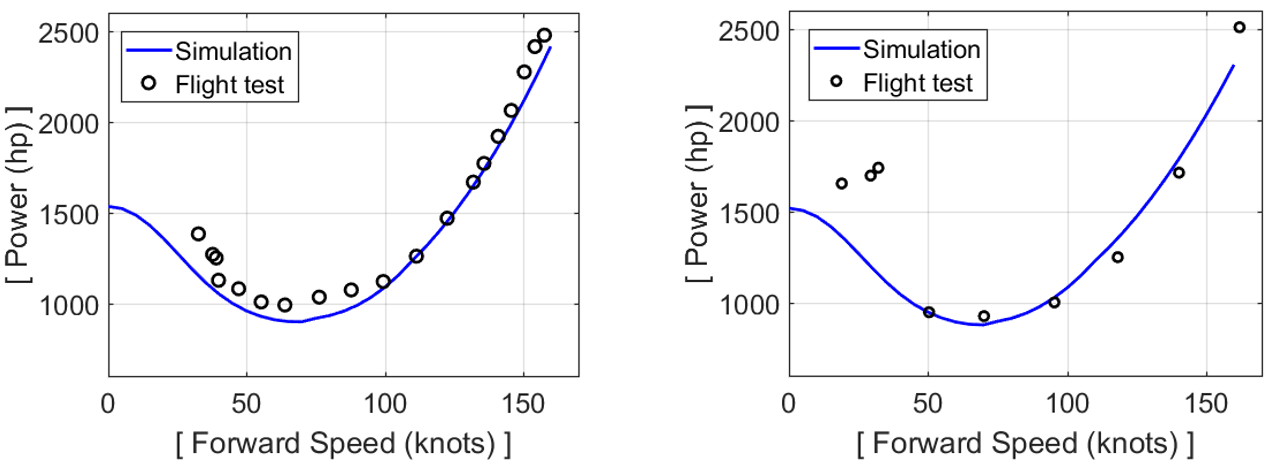}
	\caption{Main Rotor Power vs. Forward Flight Speed(Left: 16,360lbs at 3,670ft, Right: 16,000lbs at 5,250ft)}
	\label{power comp}
\end{figure}

The governing equations of a UH-60 helicopter are solved numerically to compute the trim values. The term "trim" refers to a steady flight condition where the linear accelerations along the body axes and angular accelerations about the body axes are zero. The computed trim values are compared to the test data obtained from the UH-60 flight tests conducted by the US Army \cite{abbott1982validation}. Straight and level flight is a particular case in which both the flight path angle and the rate of the turn are zero. Hover is a particular case in which the velocity is also zero. The following results presented here are simulated with a gross weight of 16,000 lbs at an altitude of 5,250 feet. First of all, the main rotor power as a function of forward flight speed is predicted and compared to flight test data in \textbf{Fig. \ref{power comp}}. Additional power comparisons are also conducted under another flight condition, which has a gross weight of 16,360 lbs at an altitude of 5,250 feet.

The comparisons show good agreements at speeds above 40 knots. At low speeds (below 30 knots), the simulated power curve under-predicts the power due to the linear inflow assumption. It could be improved by using an inflow model that captures rotor-wake interference. Typically, rotor-wake interference is stronger where the advance ratio, $\mu$ < 0.1. Thus, it requires more power at low speeds than predicted by the linear inflow models.

\begin{figure}[hbt!]
\centering
	\includegraphics[scale=0.45]{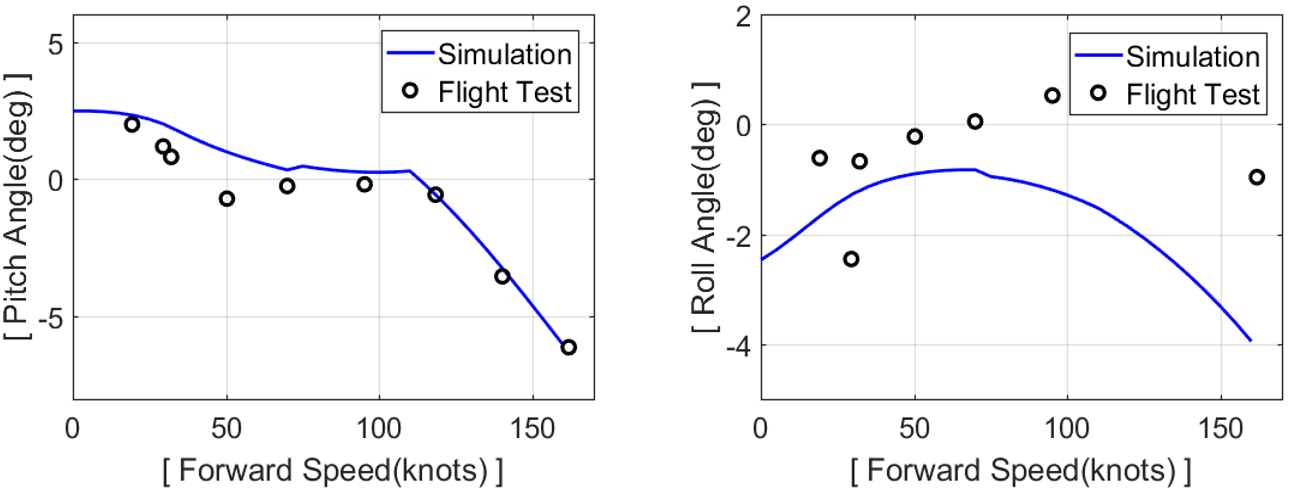}
	\caption{Fuselage Angle vs. Forward Flight Speed (16,000 lbs at 5,250 feet)}
	\label{trim angle}
\end{figure}

\textbf{Fig. \ref{trim angle}} shows the fuselage angles along the forward speed from hover to 160 knots. It is natural for the helicopter to increase the nose-down pitch angle with an increase in the forward speed. Both prediction and test data indicate that the helicopter has a nose-up or positive pitch angle at hover. Considering the angles are plotted in degrees, there is good agreement between model prediction and the flight test data.

\begin{figure}[hbt!]
\centering
	\includegraphics[scale=0.45]{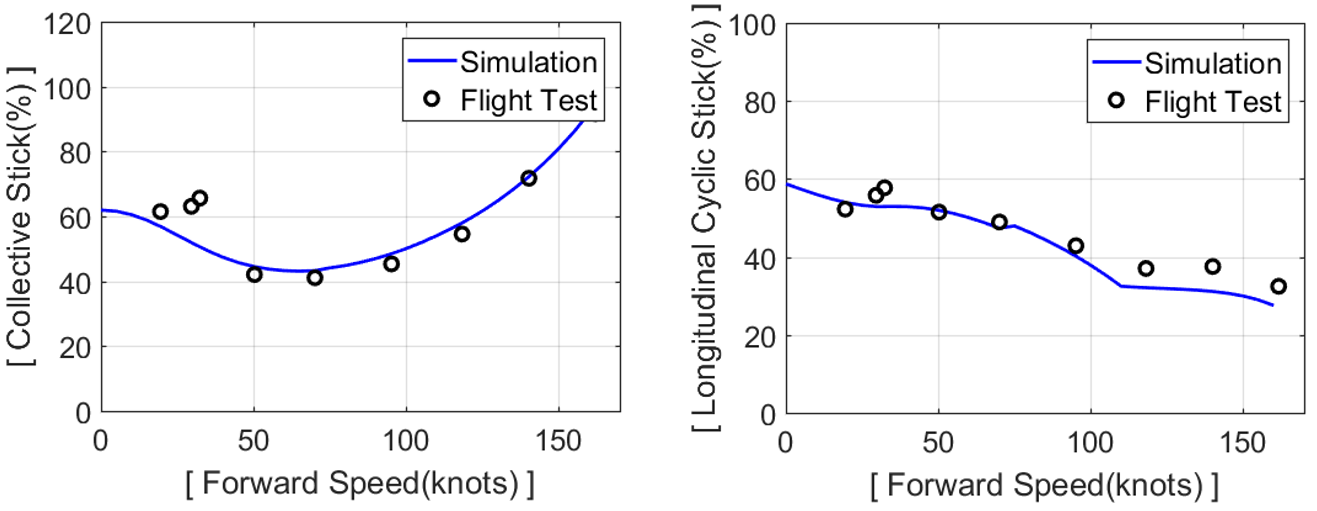}
	\includegraphics[scale=0.45]{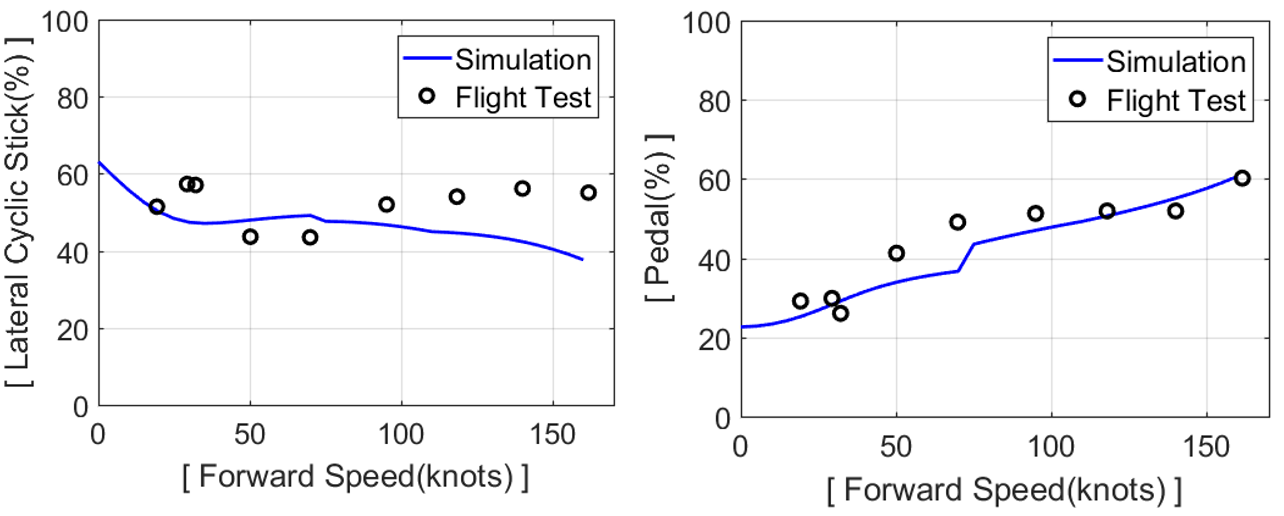}
	\caption{Control Inputs vs. Forward Flight Speed (16,000 lbs at 5,250 feet)}
	\label{controls}
\end{figure}

Control manipulations are related to flapping equations by expressing time derivatives of the first harmonic blade flapping motions as derivatives of the blade azimuth angle. These control inputs are defined as percent inputs. It means that 50$\%$ input is in a neutral position. As shown in \textbf{Fig. \ref{controls}}, the comparison results of predicted four control inputs with flight test data show good agreements.

\section{Simulation} 
\subsection{Linearized Model Extraction}
The TRAC could extract a linearized model at the desired flight conditions which can be used for various control applications. It is extracted based on a first-order Taylor series expansion of the nonlinear system governing equations about an equilibrium (trim) point. 
\begin{equation} 
f \hspace{0.4cm} + \hspace{0.4cm} \frac{\partial f}{\partial \dot{y}}\Delta \dot{y} \hspace{0.4cm} + \hspace{0.4cm} \frac{\partial f}{\partial y}\Delta y \hspace{0.4cm} + \hspace{0.4cm} \frac{\partial f}{\partial u}\Delta u \hspace{0.4cm} + \hspace{0.4cm} \cdots \hspace{0.4cm} = \hspace{0.4cm} \epsilon
\end{equation}
Accordingly, Jacobian matrices are computed at equilibrium(f = $\epsilon$ $\overset{def}{=}$ 0). 
\begin{equation}
\begin{aligned}
    E = \frac{\partial \epsilon}{\partial \dot{y}}\Big|_{trim}, \hspace{0.8cm}
    F = \frac{\partial \epsilon}{\partial y}\Big|_{trim},\hspace{0.8cm}
    G = \frac{\partial \epsilon}{\partial u}\Big|_{trim}
\end{aligned}
\end{equation}
Neglecting the higher-order terms, it yields the linearized system dynamics about equilibrium. 

\begin{equation}
\begin{aligned}
E \Delta \dot{y} \hspace{0.4cm} + \hspace{0.4cm} F \Delta y \hspace{0.4cm} + \hspace{0.4cm} G \Delta u \hspace{0.4cm} = \hspace{0.4cm} 0 
\end{aligned}
\end{equation}

By rearranging the above equation with respect to $\Delta \dot{y}$, A (stability derivatives) and B (control derivatives) matrices are computed at a given flight condition which defines the specific model (i.e. 70 knots forward flight model).
\begin{equation}
    \label{specific model}
    \begin{aligned}
    \Delta \dot{y} \hspace{0.4cm} &= \hspace{0.4cm} A \Delta y \hspace{0.4cm} + \hspace{0.4cm} B \Delta u\\
    A \hspace{0.4cm} &= \hspace{0.4cm} - E^{-1}F\\
    B \hspace{0.4cm} &= \hspace{0.4cm} - E^{-1}G
    \end{aligned}
\end{equation}

\subsection{LQR Control Design For Set-point Tracking}
Linear Quadratic Regulator (LQR) is widely used as an optimal control method \cite{bender1987linear,bemporad2002explicit}. The LQR for a set-point tracking method is used to track prescribed vehicle motions and obtain the control inputs required to fly the desired trajectory. In order to obtain the feedback gains K from the linearized dynamics, the Linear Quadratic Regulator (LQR) provides a methodology to stabilize and control a linear system by minimizing a quadratic cost function in the state deviations from targets and the control inputs. For a Linear Time-Invariant (LTI) system with dynamics given by \textbf{Eq. (\ref{specific model})}, the infinite-horizon continuous-time LQR controller yields state feedback gains K to minimize the quadratic cost function. 
\begin{equation}
\begin{aligned}
    J = \int_{0}^{\infty} (x^{T}Qx + \Delta u^{T}R \Delta u)dt\\
    (\hspace{0.2cm} \text{where} \hspace{0.3cm} x \hspace{0.3cm} = \hspace{0.3cm} y \hspace{0.3cm} - \hspace{0.3cm} y_{target} \hspace{0.2cm})\nonumber
\end{aligned}
\end{equation}
Computing the steady-state values of the states and the control inputs result in zero output error and then force them to take these values. If the desired final values of the states and control inputs are $x_{ss}$ and $u_{ss}$ respectively, then the new control formula should be 
\begin{equation}
\begin{aligned}
    \Delta u = u_{ss} - K(x - x_{ss})
\end{aligned}
\end{equation}
Plugging it in the standard form yields  
\begin{equation}
\begin{aligned}
    \dot{x} &= Ax + B(u_{ss} - Kx + Kx_{ss})\\
    y &= Cx 
\end{aligned}
\end{equation}
when x = $x_{ss}$(no error), and u = $u_{ss}$, it is expressed as
\begin{equation}
\begin{aligned}
    O &= Ax_{ss} + Bu_{ss}\\
    y_{ss} &= Cx_{ss} 
\end{aligned}
\end{equation}
It can be re-arranged in matrix form as
\begin{equation} 
            \begin{bmatrix} 
            x_{ss}\\
            u_{ss}
            \end{bmatrix}\hspace{0.5cm} = \hspace{0.5cm} 
            \begin{bmatrix} 
            A & B\\
            Cs & Ds
            \end{bmatrix}^{-1}
            \begin{bmatrix} 
            O\\
            y_{ss}
            \end{bmatrix}
\end{equation} 
In order to make it feasible, the matrix consists of A, B, Cs, and Ds components has to be invertible. Hence, Cs and Ds are selected to meet the size of the matrix. In this control system, the number of rows in Cs has to be the same as the number of control inputs, which is four. In other words, it is able to give four non-zero reference states to track and the other states are regulated to zero at the same time.
\subsection{Fully Autonomous Flight Simulation}
The helicopter ship approach and landing from 0.5 nautical miles away is used as an example for fully autonomous flight simulation. The trajectory is designed to simulate a real helicopter ship approach and landing closely. It consists of several different maneuvers which are initial descent, steady forward flight, steady coordinated turn, deceleration, and final landing. According to each flight condition, multiple different UH-60 helicopter linearized models are extracted and used. Different references are assigned to each maneuver and the LQR controller effectively regulates the error, which is the difference between the reference and current state. Gains for the controller are determined by changing weights on the Q and R matrix. Weights are carefully chosen since there is a trade-off between transient responses and control efforts. Thus, it is required to check if the control inputs are in a reasonable range. The results yield the required time to complete each maneuver and the lead/lag time for the next maneuver is determined based on the required time. The relative position of the helicopter is updated periodically by the positioning algorithm and the deviation from the reference trajectory is fed back to minimize the error and get the helicopter back on track. The fully autonomous flight simulation is conducted for 3 minutes. 
At the start of the simulation, a helicopter is flying forward with 30 knots at the height of 200 ft (60.96 m), and its initial position is defined as (0, 0, -60.96) in the earth-fixed frame. A target ship is moving forward with 10 knots at (679.7285, -88, -5) in the earth-fixed frame. All units are in meters and in order to demonstrate the trajectory more intuitively, the sign of the Z-axis component in the earth-fixed frame is reversed in the following plots (positive upwards). In the entire trajectory plots, the red marker represents the ship trajectory and the blue line represents the helicopter trajectory.
\begin{figure}[hbt!]
\centering
   \begin{minipage}{0.48\textwidth}
     \centering
     \includegraphics[scale=0.26]{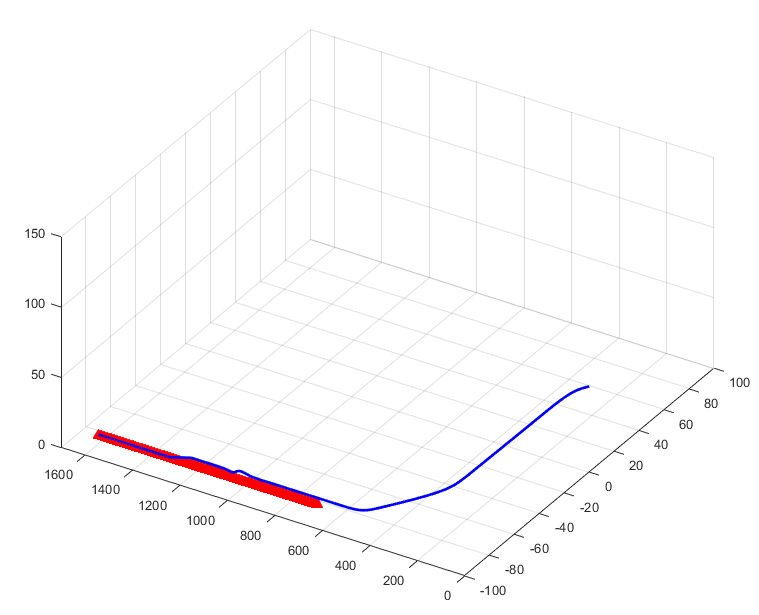}
     \caption{Entire Trajectory in diagonal view}
   \end{minipage}
   \begin {minipage}{0.48\textwidth}
     \centering
     \includegraphics[scale=0.26]{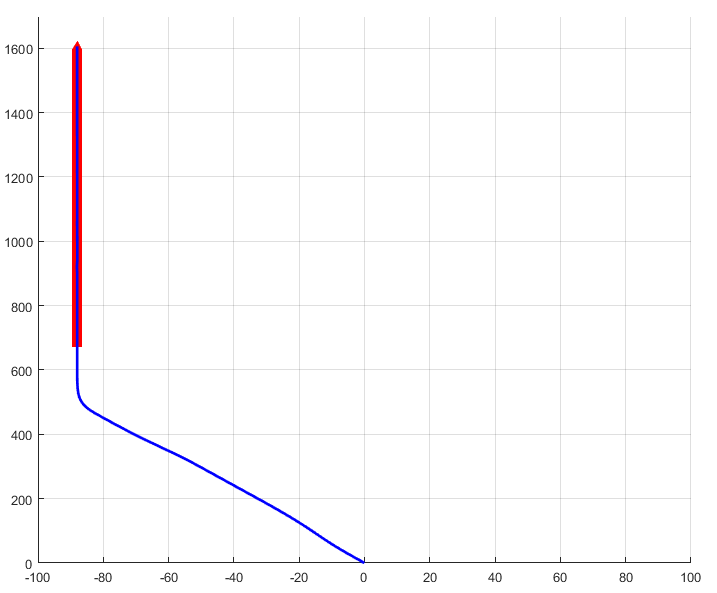}
     \caption{Entire Trajectory in top view}
   \end{minipage}
   \begin{minipage}{0.48\textwidth}
     \centering
     \includegraphics[scale=0.26]{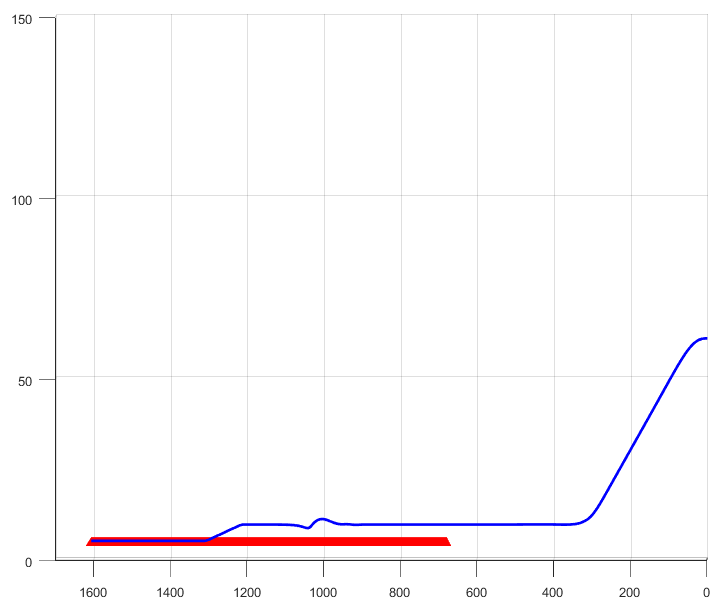}
     \caption{Entire Trajectory in side view}
   \end{minipage}
   \begin {minipage}{0.48\textwidth}
     \centering
     \includegraphics[scale=0.26]{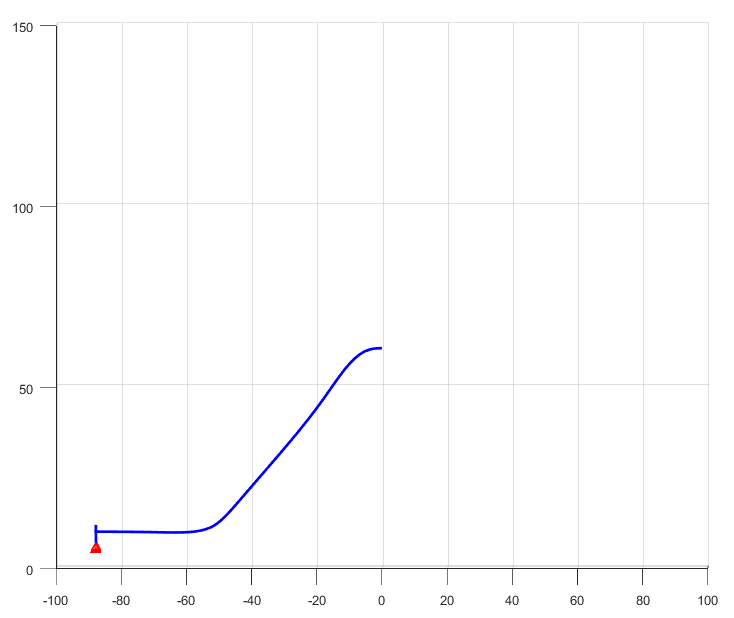}
     \caption{Entire Trajectory in rear view}
   \end{minipage}
\end{figure}
\\

In order to check the trajectory more specifically, relative distance in the earth-fixed frame is investigated and the final landing position is also plotted. The relative distance is calculated by the "Ship position" - "Helicopter position" in the earth-fixed frame. $\Delta X$, $\Delta Y$, and $\Delta Z$ are the relative distance along the earth-fixed X, Y, Z-axis. In the following plots, the Z-axis component sign also follows a re-defined direction (positive upwards). The final landing circle on a flight deck means that landing anywhere inside the circle is safe. Thus, it can be considered as an allowable error range. Final values of $\Delta X$, $\Delta Y$ are considered an error, but  $\Delta Z$ is the summation of the distance from the landing gear to the CG (48.26 cm) and error. Hence, the final errors mean the deviation from the center of the circle and are expressed along each axis in meters (0.0353, 0.0728, 0.0037). Considering the size of the circular boundary, this is a good enough result.

\begin{figure}[hbt!]
\centering
   \begin{minipage}{0.48\textwidth}
     \centering
     \includegraphics[scale=0.5]{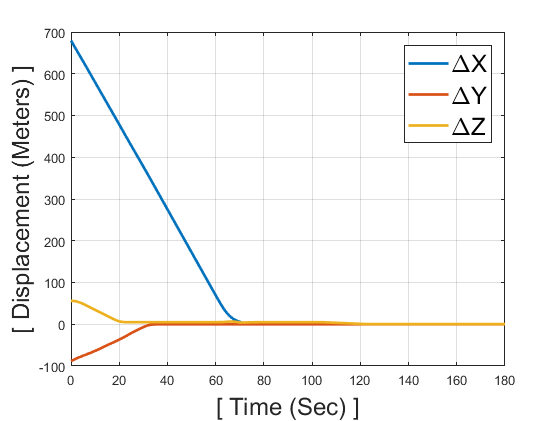}
     \caption{Relative Displacement in time}
   \end{minipage}
   \begin {minipage}{0.48\textwidth}
     \centering
     \includegraphics[scale=0.5]{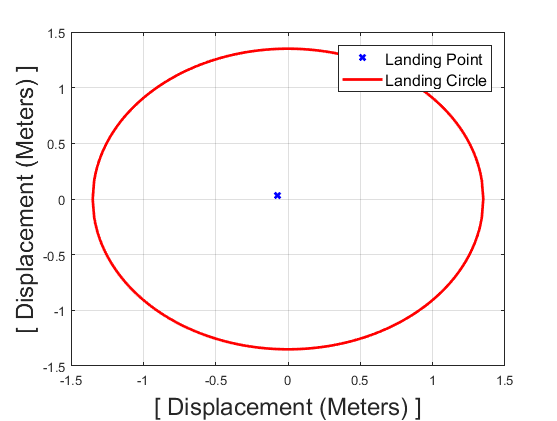}
     \caption{Final Landing Point}
   \end{minipage}
\end{figure}
The MATLAB simulation results are visualized by using the X-Plane 11 flight simulator software due to its excellent graphics quality and capability to reconstruct the flight by using a flight data recorder (FDR) file. The FDR is a device to collect and record data from aircraft sensors and therefore, it is commonly used for accident investigation. For this visualization, the results are re-written in an FDR format that is loaded into X-Plane. It visualizes the complete helicopter behavior by taking those values directly from the simulation results. It visualizes the entire helicopter ship landing maneuver. The simulation videos can be seen here: \href{https://youtu.be/1Lez4s4HhWM}{cockpit view}, \href{https://youtu.be/8YW5av_MeW8}{ship view}.

\begin{figure}[hbt!]
\centering
   \begin{minipage}{0.48\textwidth}
     \centering
     \includegraphics[scale=0.235]{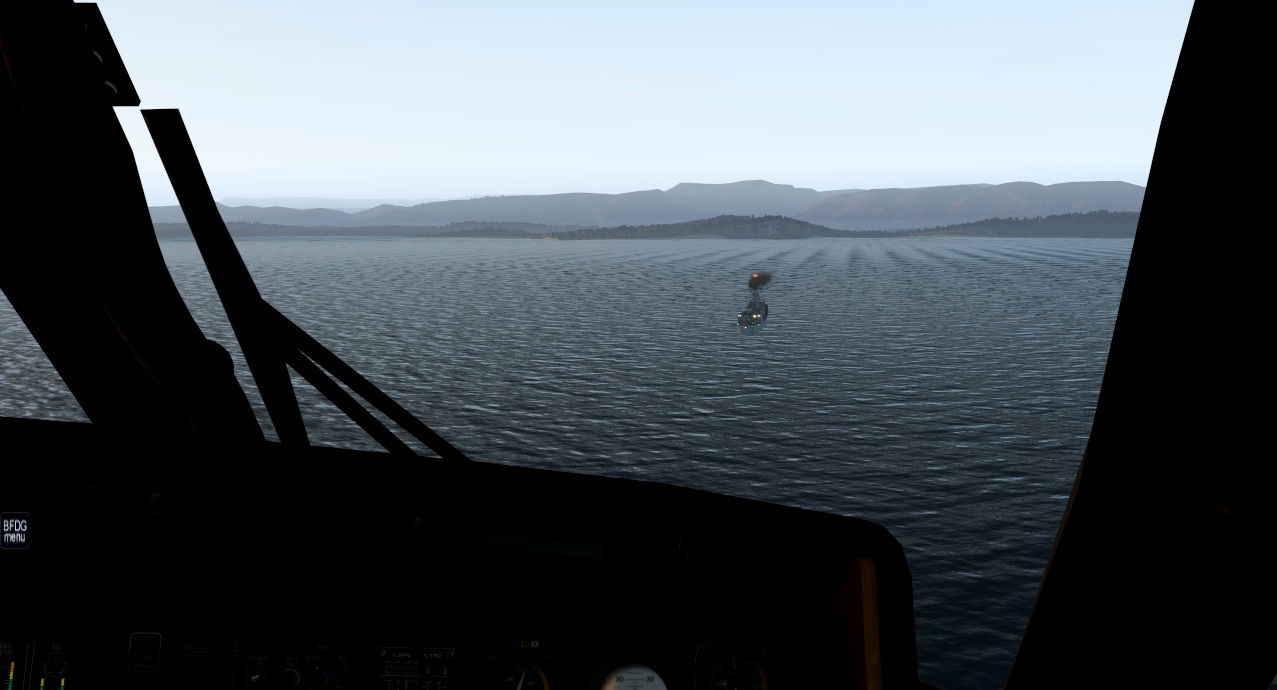}
     \caption{\href{https://youtu.be/1Lez4s4HhWM}{Cockpit View at Initial Position}}
   \end{minipage}
   \begin {minipage}{0.48\textwidth}
     \centering
     \includegraphics[scale=0.23]{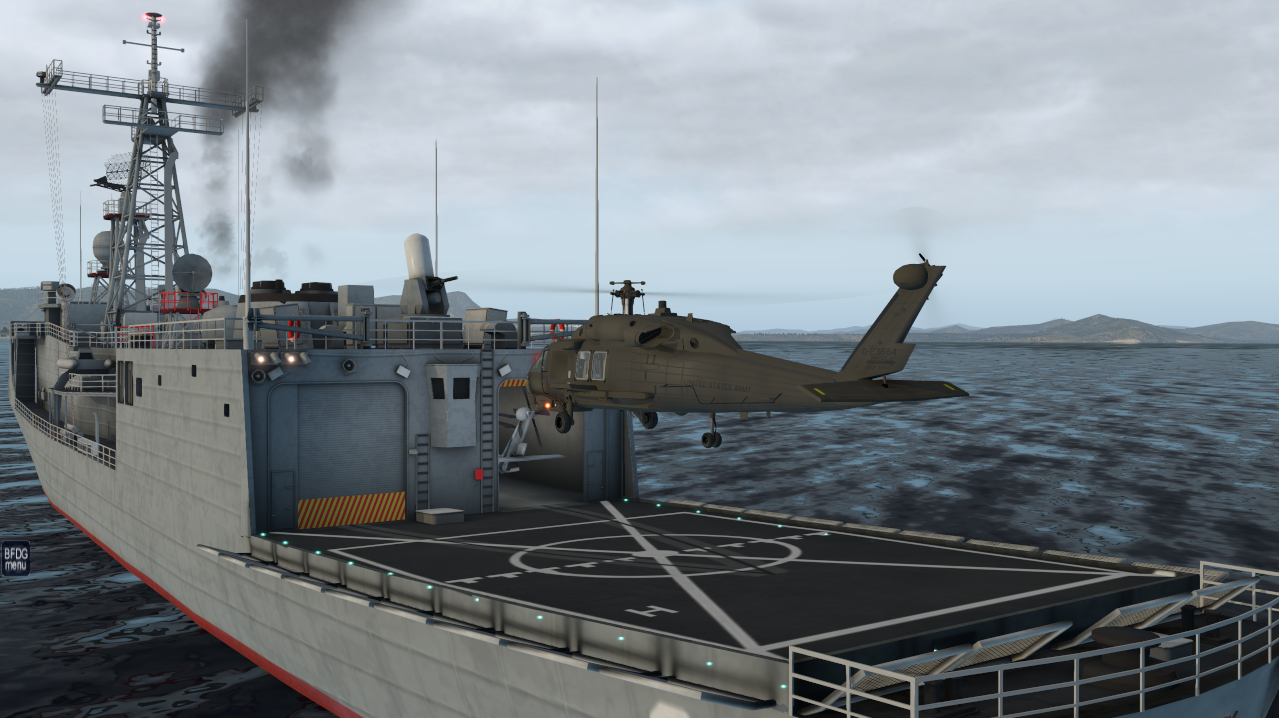}
     \caption{\href{https://youtu.be/8YW5av_MeW8}{External View at flight deck}}
   \end{minipage}
\end{figure}

\section{Summary}
TRAC has been developed as a comprehensive helicopter flight dynamics tool to quickly predict the impact of newly designed components and/or design changes on flight performance. Thus, dynamic/aerodynamic models of each component of a representative helicopter (UH-60) have been developed individually and integrated as a complete helicopter model. It has adopted wind-tunnel data and empirical data to improve prediction results. The prediction results including helicopter power, attitude, rotor blade angles, and control inputs have been validated with the US Army UH-60 flight test data.\\
\indent The linearized models are extracted at various flight conditions and used for achieving a fully autonomous flight by the LQR controller. Since the LQR method requires the constructed matrix to be invertible, the number of references has to be the same as the number of control inputs. Hence, four references are uniquely selected in each phase of flight for effective trajectory tracking. A fully autonomous flight is simulated from approach to landing on a ship and it is visualized by using X-Plane flight simulator program.\\
\indent TRAC is a complete software package which encompasses modeling, trim analysis, and autonomous flight simulation. Even though the UH-60 helicopter is modeled currently, TRAC could be configured for a new type of helicopter by adding or re-configuring component models and the performance can be predicted. Furthermore, any type of autonomous flight can be implemented by using the featured LQR method or any other linear controller since it extracts linearized models at the desired flight conditions.

\section*{Appendix}
\vspace{0.1cm}
\centering UH-60 Helicopter Configuration
\begin{table}[hbt!]
	\centering
	\begin{tabular}{|lll|} \hline
    Main Rotor && \\ \hline
    Number of blades && 4  \\  
    Radius R, ft && 26.83  \\  
    Blade chord c, ft && 1.75  \\  
    Rotational speed, rad/sec && 27.0  \\  
    Tip speed, ft/sec && 724.41  \\  
    Longitudinal mast tilt, deg && -3.0  \\  
    Airfoil section && SC 1095  \\  
    First airfoil section, ft && 5.08  \\  
    Blade precone, deg && 0.0  \\  
    Linear blade twist, deg && -18.0  \\  
    Solidity && 0.083  \\  
    Lock number && 5.11  \\  
    Control phase shift && -9.7  \\  \hline
    Tail Rotor && \\ \hline
    Number of blades && 4  \\  
    Radius, ft && 5.5  \\  
    Blade chord, ft && 0.81  \\  
    Rotational speed, rad/sec && 124.62  \\  
    Tip speed, ft/sec && 685.41  \\  
    Rotor shaft cant angle, deg && 20.0  \\   \hline
    Fuselage && \\ \hline
    Gross weight, lbs && 16000.00  \\  
    Pitch inertia $I_{yy}$, lbs$\cdot ft^{2}$ && 38512.0  \\  
    Roll inertia $I_{xx}$, lbs$\cdot ft^{2}$&& 4659.0  \\  
    Yaw inertia $I_{zz}$, lbs$\cdot ft^{2}$&& 36796.0  \\  
    $I_{xz}$, lbs$\cdot ft^{2}$ && 1882.0  \\  
    Horizontal tail surface area, $ft^{2}$ && 45.00  \\   \hline
	\end{tabular}
	\caption{Main Parameters of the UH-60 helicopter configuration}
	\label{config table}
\end{table}

\section*{Acknowledgment}
The author's graduate study was partially supported by a fellowship from the Republic of Korea Navy.

\bibliography{sample}

\end{document}